\documentclass[10pt,letterpaper]{article}
\usepackage[top=0.85in,left=2.75in,footskip=0.75in,marginparwidth=2in]{geometry}

\usepackage[utf8]{inputenc}

\usepackage{cite}

\usepackage{nameref,hyperref}

\usepackage[right]{lineno}

\usepackage{microtype}
\DisableLigatures[f]{encoding = *, family = * }

\raggedright
\usepackage{changepage}

\usepackage[aboveskip=1pt,labelfont=bf,labelsep=period,singlelinecheck=off]{caption}

\makeatletter
\renewcommand{\@biblabel}[1]{\quad#1.}
\makeatother

\usepackage{lastpage,fancyhdr,graphicx}
\usepackage{epstopdf}
\fancyheadoffset[L]{2.25in}
\fancyfootoffset[L]{2.25in}

\usepackage{color}

\definecolor{Gray}{gray}{.25}

\usepackage{graphicx}

\usepackage{sidecap}

\usepackage{wrapfig}
\usepackage[pscoord]{eso-pic}
\usepackage[fulladjust]{marginnote}
\reversemarginpar

\begin{document}
\vspace*{0.35in}

\begin{flushleft}
{\Large
\textbf\newline{A random version of principal component analysis in data clustering }
}
\newline
\\
Luigi Leonardo Palese\textsuperscript{1,*},
\\
\bigskip
\bf{1} University of Bari "Aldo Moro", Department of Basic Medical Sciences, Neurosciences and Sense Organs (SMBNOS), Bari, 70124, Italy
\\
\bigskip
* luigileonardo.palese@uniba.it

\end{flushleft}

\section*{Abstract}
Principal component analysis (PCA) is a widespread technique for data analysis that relies on the covariance-correlation matrix of the analyzed data. However to properly work with high-dimensional data, PCA poses severe mathematical constraints on the minimum number of different replicates or samples that must be included in the analysis. Here we show that a modified algorithm works not only on well dimensioned datasets, but also on degenerated ones.


\section*{Introduction}
Science today is surrounded by large amounts of data. These are produced by techniques and instruments able to measure a huge number of variables on a large number of samples, they are deposited in an increasing number of online databases that grow exponentially, and also modern numerical simulations can produce very large and high-dimensional outputs. The challenge of the growing size of data concerns all sciences, but the field in which we have seen the most spectacular growth is probably that of life sciences, where the advancement of genomics, proteomics and high-throughput technologies has produced an overwhelming amount of data, more and more often freely available to all researchers. Beside the large number of samples, these data are big also because they are high-dimensional: this means that each sample, or instance, of a typical dataset contains a large number of degree of freedom. Such high-dimensionality makes visualization and exploration of samples and datasets very difficult. To overcome these limitations, a series of techniques have been developed that help researchers in visualization, exploration and mining of large data. \cite{van2009}

Among the various algorithms that reduce the dimensionality of data, while retaining the important information, one of the most successful is principal component analysis (PCA). \cite{ringner2008} PCA has been reinvented several times, but it has been developed in its modern form by Pearson and Hotelling. \cite{pearson1901, hotelling1933,bro2014} How PCA works is recalled in the Methods section, but here it is important to note that, in its classical implementation, PCA relies on the covariance (or also correlation) matrix of the analyzed data. This is actually a point often overlooked by end-users, but it should be stressed that the number of samples needed to accurately estimate the covariance/correlation matrix of a system containing $n$ degree of freedom should be (much) larger than $n$. Otherwise the covariance/correlation matrix will be full of spurious correlations, as well as rank deficient from a mathematical point of view if the number of samples is less than $n$. However here we will show that what is important for the functioning of the method it is not a particular covariance/correlation matrix, but rather the symmetry that characterizes this type of matrices.

\section*{Results and Discussion}
We developed the method in order to calculate the PCA of a set of protein crystallographic structures. Proteins are structurally and dynamically complex objects. \cite{frauenfelder2002,palese2013cplx} Because molecular dynamics (MD) is actually at a level of accuracy \cite{dror2012} that permits to predict experimentally observables, \cite{bossis2011} it is nowadays a standard tool for the dynamical characterization of proteins. In the analysis of MD trajectories PCA is of widespread use: \cite{kitao1999} the high-dimensional large number of different molecular conformations that constitute the output of a MD experiment is an ideal dataset for PCA. \cite{bossis2013} On the other hand, the number of protein structures reported in the Protein Data Bank \cite{berman2000} (PDB) is collectively large, but there are few structures of a single protein. Although it is possible to find dozens or even hundreds of versions of a single protein in the PDB, the number of available structures is incomparably smaller than the number of degree of freedom of a protein. So while PCA can be used in the analysis of the thousands of conformations obtained from MD simulations, in its classical implementation PCA can not be used in the analysis of the experimental structures as the low number of different conformations reported in the PDB does not allow an accurate calculation of the covariance matrix. To perform a structural analysis similar to the PCA we choose to analyze the human serum albumin (HSA) available structures in the PDB. HSA, \cite{fanali2012} the most abundant protein in plasma, is a monomeric multi-domain molecule. HSA is a non-glycosylated, all-$\alpha$ protein chain of 65 kDa, with a globular heart-shaped conformation consisting of three homologous domains (I-III). Each domain is composed by two subdomains (A and B). It is an important transport protein with different binding sites able to accommodate a number of chemically different ligands. HSA represents the main carrier for fatty acids (there are seven binding sites for fatty acids, labeled as FA1 to FA7), and it is a depot and carrier for exogenous compounds (mainly, but not exclusively at the Sudlow's sites I and II), thus affecting the pharmacokinetics of many drugs.

Among the available structures, we selected 58 structure for the analysis (see Methods for selection criteria). After structural alignment, the $\alpha$-carbon atom Cartesian coordinates were extracted and arranged in a data matrix in which each row represented a single HSA structure. Thus the data matrix was composed of 58 rows and 1695 columns (565 $\alpha$-carbon atoms were finally included in the analysis; see the Methods section). This is clearly a degenerated dataset, as it is impossible to obtain the true correlation matrix of a multivariate system with 1695 degree of freedom by using only 58 samples. If we calculate the correlation matrix, this will be, at best, only a rank deficient approximation of the true one in which a large number of false correlations must be expected. While it is true that, using a careful error handling (and silencing) program, or also using algorithms that estimate the principal components without ever computing the covariance matrix, \cite{roweis1998} it is generally possible to calculate the first principal components, the classical PCA is not calculable on this dataset.

It should be considered that what really we are interested in is not the identification of the axes that describe the greatest variance of the data (axes which do not have a particular {\it{a priori}} meaning), but instead an orthogonal linear transformation of data that could be useful in exploratory data analysis. We can relax the request that the correlation-covariance matrix (the true or the approximated one) is needed for such transformation: it is possible that what is important in PCA as clustering tool may not be the use of a {\it{particular}} matrix, but instead of a matrix belonging to a particular {\it{symmetry class}}. The bases for such an hypothesis are rooted in the fact that good models for the covariance matrices for the protein configurations obtained from MD \cite{palese2015bpc,palese2015jpcb,palese2016jpcb} are a class of symmetric random matrices. \cite{edelman2013} Moreover, the consequences of the Johnson-Lindenstrauss lemma, \cite{johnson1984} and the fact that in the Pearson original view \cite{pearson1901,bro2014} of PCA which is important is the subspace and not the axes as such, furnish us a further justification. Thus we applied to the albumin dataset a variant of the PCA algorithm, in which a square symmetric random matrix was used, instead of the correlation-covariance one (this matrix was obviously of dimension of 1695). We will refer to this algorithm as random component analysis (RCA). The detailed algorithm is described in the Methods section, and an easily customizable implementation is reported in Supplementary Information section.

\begin{figure}[ht]
\centering
\includegraphics[width=0.5\linewidth]{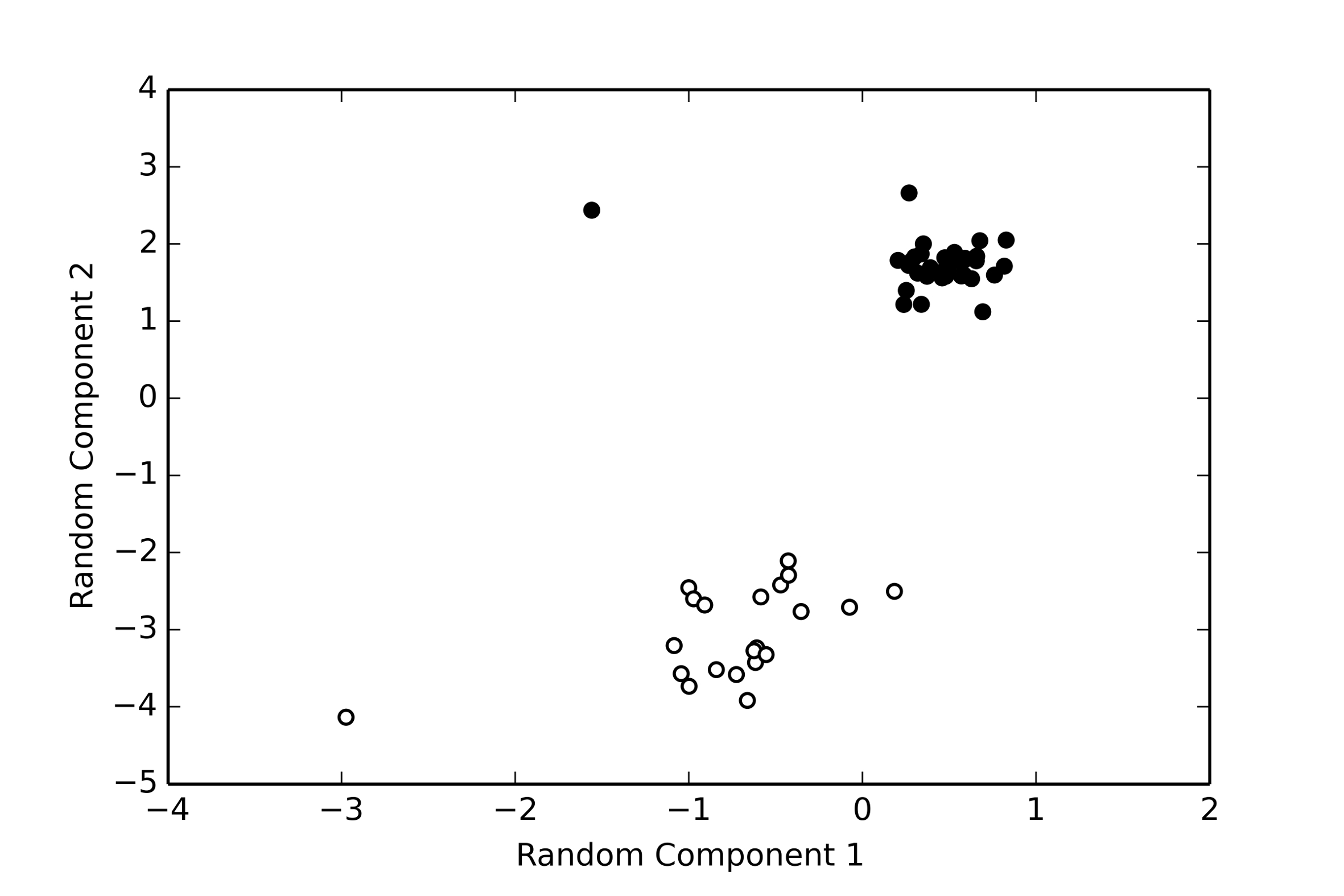}
\caption{\color{Gray} \textbf{Random component analysis of the HSA structures.} The Figure reports a random component analysis on the HSA structures contained in the dataset described in the text. The HSA structures with bound fatty acids are reported as solid (black) circles, whereas the structures without bound fatty acids are reported as void (white) circles. The algorithm clearly permits to differentiate two clusters of structures in the dataset, and the discriminant is the presence of absence, respectively, of bound fatty acids. Two similar cluster of structures have been obtained in all the random component analysis calculations carried out on the HSA dataset (see Supplementary Information).}
\label{fig:figure1}
\end{figure}

The results of this analysis are reported in Figure \ref{fig:figure1}. As can be easily appreciated by inspecting the figure, RCA leads to two well defined clusters of structures, and what is more interesting is that one cluster contains all and only the HSA molecules with bound fatty acid, the other one only structures without fatty acid. 
These cluster are reproducible (see Supplementary Figure 1) and are similar to those obtained by different protocols (see Methods and Supplementary Figure 2). It worth noting that a large number of structural and functional works on HSA lead to the conclusion that two structures, possibly related to the presence of fatty acids, are discernible for this protein. \cite{fanali2012,ascenzi2010} Our RCA analysis permits to go further, as it clearly demonstrates that the only discriminant for such structural switch in the whole dataset is the presence or absence of bound fatty acid.

\begin{figure}[ht]
\centering
\includegraphics[width=\linewidth]{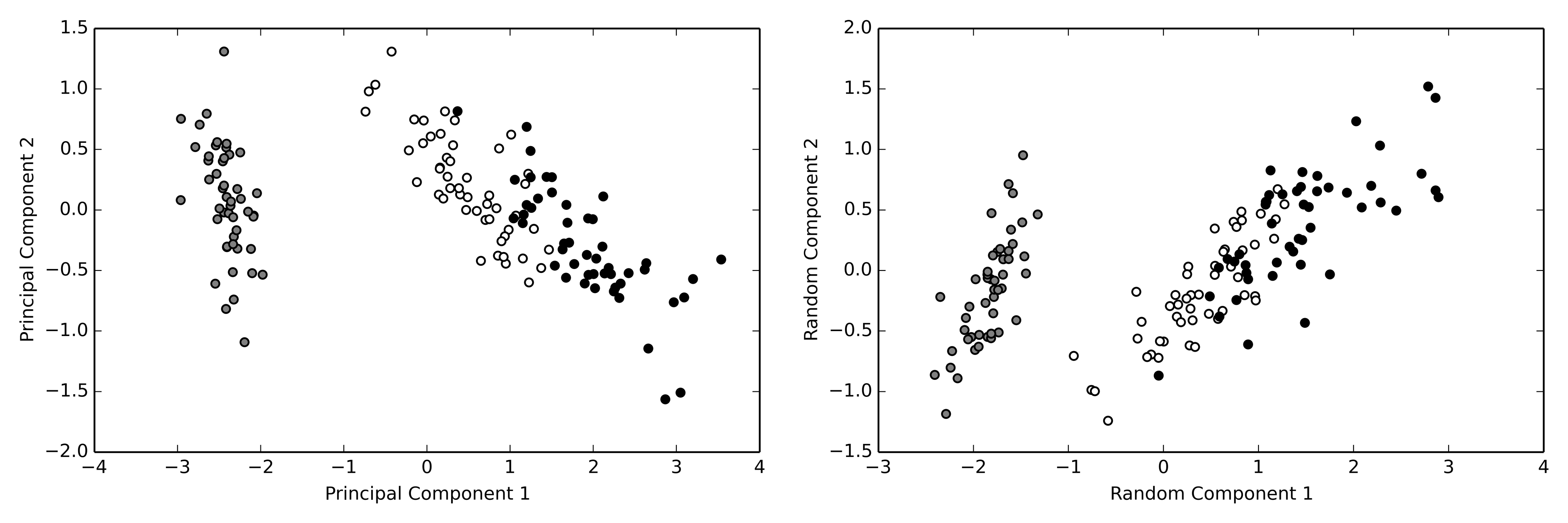}
\caption{\color{Gray} \textbf{The Iris dataset.} Principal component analysis (left) and random component analysis (right) of the Iris dataset are reported. The Iris dataset is a simple but classical benchmark for the clustering algorithms. This dataset contains 150 entries, 50 for each of the species {\it{Iris virginica}} (black), {\it{Iris setosa}} (gray) and {\it{Iris versicolor}}  (white). Both algorithms easily differentiate the {\it{Iris setosa}} cluster, whereas the other species can be only partially discriminated by the algorithms of this class. The full set of random component analysis carried out on the Iris dataset shows similar clustering results and it is reported in Supplementary Information.}
\label{fig:figure2}
\end{figure}

While RCA has been developed for degenerated datasets (that means datasets that are characterized by a larger number of degree of freedom respect to the number of samples that can be analyzed) we tested it also on well sized datasets. These were retrieved the from the UCI (University of California at Irvine,  School of Information and Computer Science) Machine Learning Repository (Lichman, M., http://archive.ics.uci.edu/ml). The results of PCA and RCA on the classical Iris dataset are reported in Figure \ref{fig:figure2}. 
It can be stated that RCA is at least not inferior to PCA in clustering purposes, and the results are reproducible, as reported in Supplementary Figure 3. Similar results have been obtained with the Wine dataset, which are reported in the Supplementary Figure 4.

\section*{Conclusions}
The algorithm proposed in this communication is easy to implement, conceptually simple and numerically robust. It is another example of useful application of random matrix theory, \cite{palese2015bpc,palese2015jpcb,palese2016jpcb,edelman2013} whose pervasiveness is even more evident in a large number of fields. This work demonstrates that what is important for clustering efficiency of PCA is not the exact form of the covariance-correlation matrix, but instead simply its symmetry, as in our RCA algorithm. The fact that good and informative clustering can be achieved by random projection is nowadays an emerging concept that, beside practical applications, could have far reaching implications also from a conceptual point of view. Finally, this work suggests that an excessive confidence on correlations (which are often spurious) and on large covariance should be avoided, if a simple random matrix could well surrogate them in cluster generation. 

\section*{Methods}
\subsection*{HSA database construction}
In order to build up a suitably large dataset of structures we searched in the Protein Data Bank \cite{berman2000} (www.rcsb.org) for the albumin structures, with the constraints of specie (human), single protein type in the structure, and resolution of 3.30~{\AA}  or better. After this initial screening, because some N- and C-terminal residues are often not present in the deposited structure, and in order to include the largest possible number of structures as complete as possible, the ones starting after the SER 5 and ending before ALA 569 were excluded from the database. Finally, the structures containing a number of $\alpha$-carbon atoms different of 565 were also excluded. The final dataset contained 58 structures \cite{sugio1999, bhattacharya2000jbc, bhattacharya2000jmb, petitpas2001jmb, petitpas2001jbc, petitpas2003pnas, wardell2002, zunszain2003, he1992, ghuman2005, yang2007, ryan2011, zhu2008, guo2009, hein2010, buttar2010, he2011, sivertsen2014, wang2013jbc, wang2013bba, zhang2015structural, bijelic2016} which are reported in the Supplementary Table 1. 

A pdb file for each of these structures has been written in VMD \cite{humphrey1996} (from SER 5 to ALA 569); these structures were aligned using MultiSeq \cite{roberts2006} and the pdb files were updated to the new coordinates. The same software was used to calculate the distance trees (RMSD and Qh style). \cite{odonoghue2005,russell1992} The obtained tree are reported as Supplementary Figure 2 (see also the Supplementary Table 2). 

To obtain the dataset in a matrix form, the pdb files were loaded in VMD and the $\alpha$-carbon atom coordinates were extracted and written in a text file such that each row described a structure, by a Tcl (www.tcl.tk) script. Curly brackets in the raw text file were eliminated by vim scripting (www.vim.org), so as to obtain the data matrix in a readable file format by the numerical analysis software. 

\subsection*{The PCA and RCA algorithms}
PCA was based on the eigenvector decomposition of the correlation matrix. \cite{van2009,ringner2008,bro2014,bossis2013,palese2015bpc,palese2015jpcb,shlens2014tutorial} After the centroid subtraction, the covariance matrix of the dataset matrix described above was obtained as 
\[
C_{ij} = \langle (x_{i} - \langle x_{i} \rangle ) (x_{j} - \langle x_{j} \rangle) \rangle
\]
where $\langle \dots \rangle$ represents the average over all the conformations in the dataset. The correlation matrix was calculated from this matrix as 
\[
P_{ij} = \frac{C_{ij}}{\sqrt{C_{ii} C_{jj}}}
\]
with obvious meaning of symbols. This square symmetric matrix was diagonalized 
\[
R^{T} P R = \Lambda
\]
using standard numerical routines (see below), where $R$ is an orthonormal transformation matrix, the superscript $^T$  means transposition and $\Lambda$ is a diagonal matrix whose elements are the eigenvalues. The empirical matrix was projected onto the eigenvectors to give the principal components. 

The RCA was performed exactly as the PCA, except for the fact that the square symmetric correlation matrix was replaced by a random symmetric matrix, obtained as
\[
M = \frac{G + G^{T}}{2}
\]
where $G$ was a normal distributed random square matrix. So this algorithm could be conceived as a version of classical PCA with relaxed constraints respect to the matrix to be used in calculating the new  orthonormal reference system, where only the matrix symmetry is preserved.

\subsection*{Software implementation and code availability}
The PCA and RCA algorithms were implemented in the Python language (www.python.org) in an IPython notebook. \cite{perez2007} The NumPy numerical software library \cite{vandervalt2011} was used, which is part of the Scipy \cite{oliphant2007} software package. The Pandas \cite{mckinney2010} and Matplotlib \cite{hunter2007} packages were used to import the Iris and Wine datasets and to obtain the all graphical outputs, respectively (both packages were obtained from Scipy; www.scipy.org). The implementation of these algorithms is reported in Supplementary Information in Python format. Two versions of the RCA algorithm are reported: the first one requires the dataset and the dimension of the dummy correlation matrix as arguments, while the second requires as arguments the dataset and the random matrix that will be used for the calculation of the orthogonal projection system. This last function could be useful if one would save a particularly interesting matrix for further analysis. These files are easily customizable; as it is provided, the software requires (very) few seconds for the download and analysis of the proposed datasets (the HSA dataset described above, the Iris and Wine datasets) on an Intel Core i7 machine or a Xeon equipped workstation, both running Ubuntu 14.04 LTS. Very large datasets (as in the case of MD outputs; not shown) could require up to (also several) minutes to be analyzed. Because the RCA algorithm performs a random projection it is preferable to carry out multiple runs of it. In a small percentage of cases (no more than 5\% - 10\% of the tests used for this work) the algorithm does not get a projection that separates the samples in different clusters, and this is the only drawback of the simple implementation of the RCA algorithm here described. 


\bibliography{sample}

\bibliographystyle{abbrv}
\clearpage
\section*{Supplementary Information}
Supplementary figures and code
\newpage
\begin{figure}
\centering
\includegraphics[width=\linewidth]{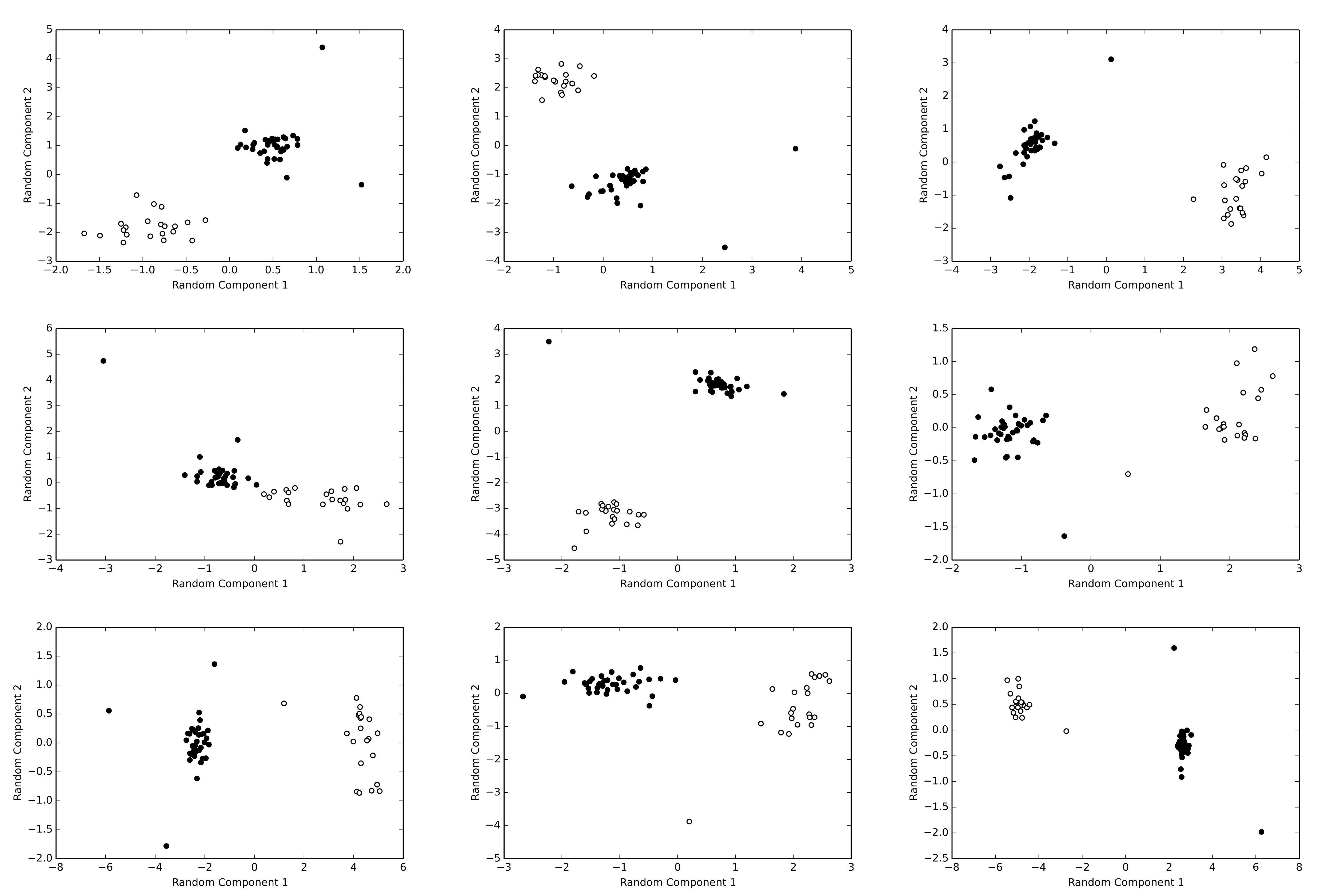}
\caption*{{\bf Supplementary Figure 1. HSA random component analysis.} The figure reports 9 consecutive application runs of the random component analysis algorithm on the HSA dataset described in the main text. The HSA structures with bound fatty acids are reported as black circles, whereas the HSA structures without fatty acids (either ligand free or with chemically different ligands) are reported as open (white) circles. It is evident that the two class of HSA structures are in different clusters.}
\end{figure}

\clearpage
\begin{figure}
\centering
\includegraphics[width=\linewidth]{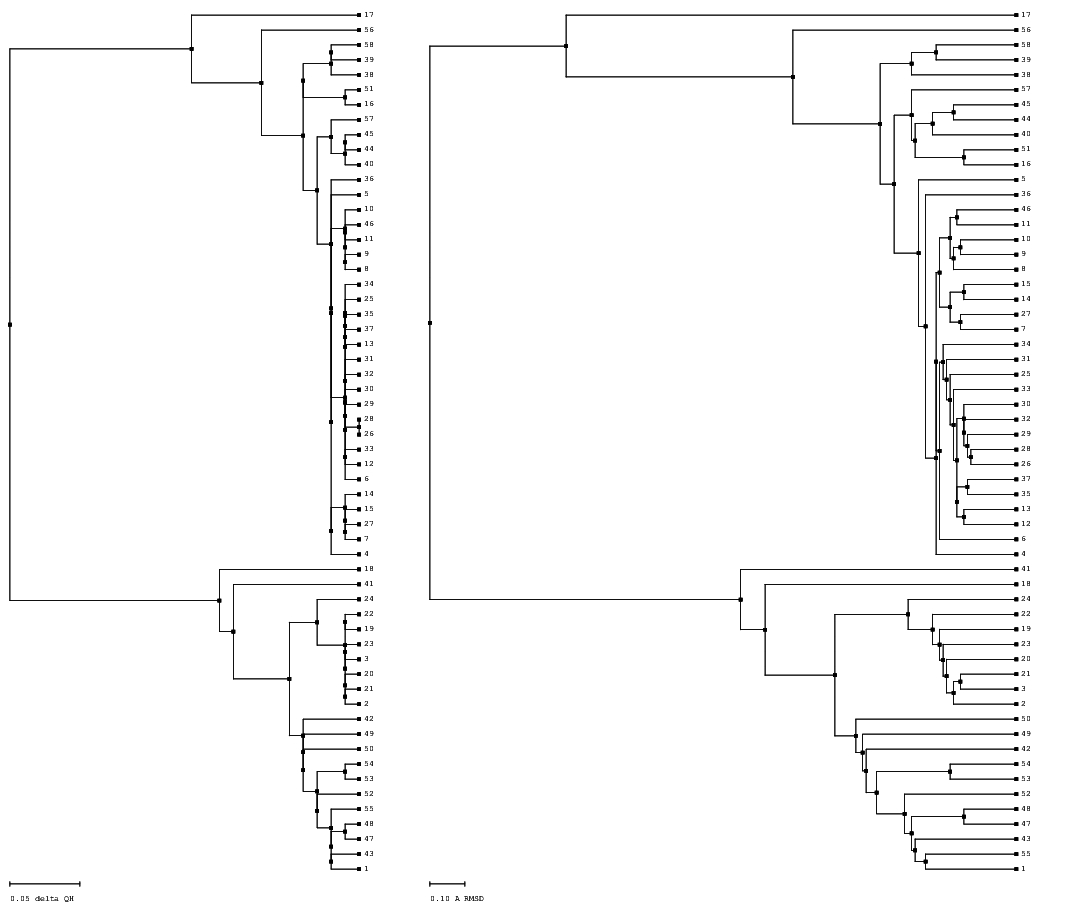}
\caption*{{\bf Supplementary Figure 2. Tree clustering of HSA dataset.} The Qh (left) and RMSD (right) trees have been obtained with the MultiSeq program as described in the main text. The numbers correspond to the protein PDB entries as reported in the Supplementary Table 1 (see below). Both algorithms recognize two clusters (cluster A and B in Supplementary Table 2): cluster A contains the HSA structures without bound fatty acids, while cluster B contains HSA molecules with bound fatty acid.}
\end{figure}

\clearpage

\begin{figure}
\centering
\includegraphics[width=\linewidth]{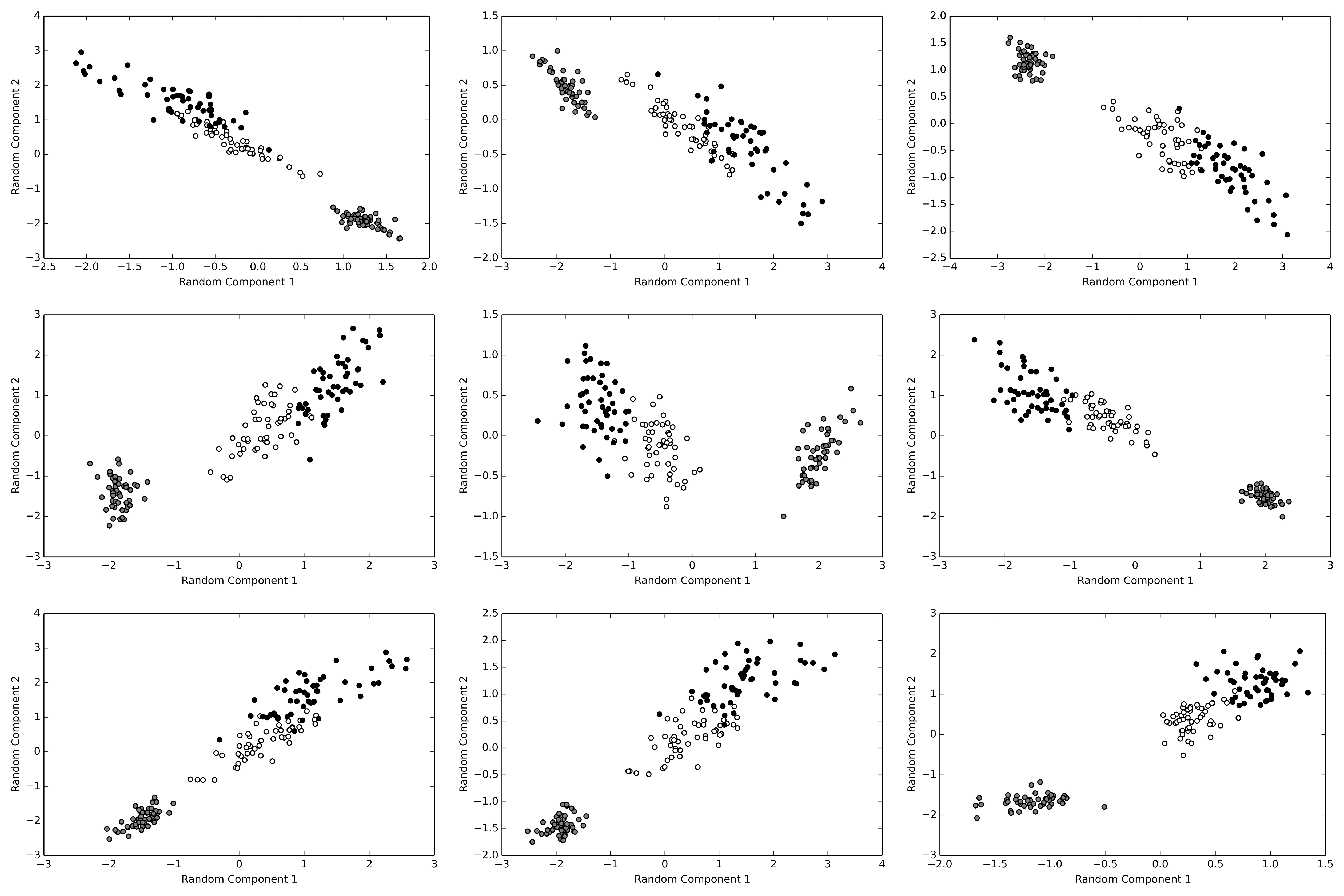}
\caption*{{\bf Supplementary Figure 3. Random component analysis of Iris dataset.} The figure reports 9 consecutive application runs of the random component analysis algorithm on the Iris dataset. The {\it{Iris virginica}} specie is reported in black, the {\it{Iris setosa}} in gray and {\it{Iris versicolor}} in white. As in the case of PCA, the algorithm easily differentiate the {\it{Iris setosa}} cluster, whereas the other species can be only partially discriminated by the algorithms of this class (linear classifiers).}
\end{figure}

\clearpage

\begin{figure}
\centering
\includegraphics[width=\linewidth]{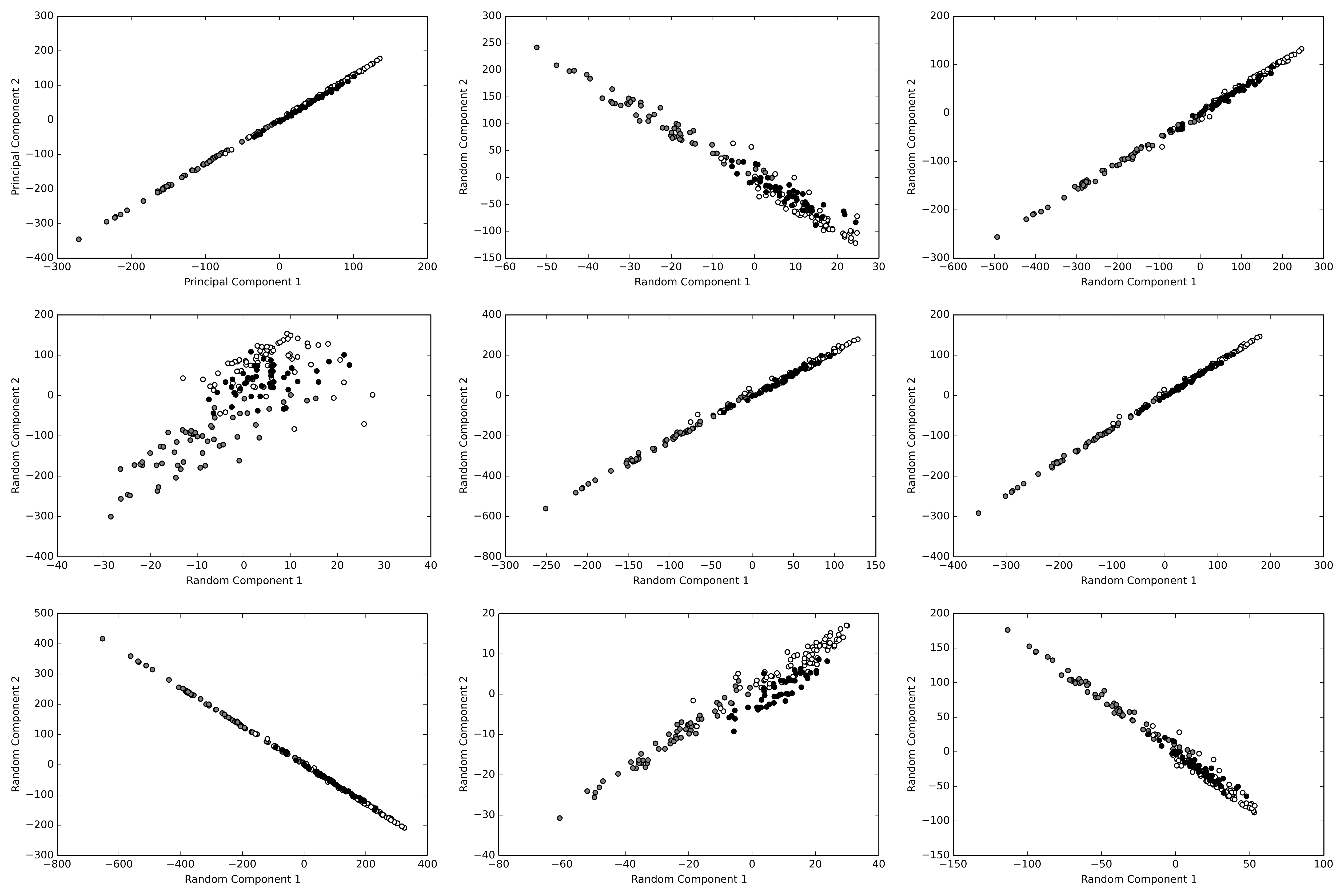}
\caption*{{\bf Supplementary Figure 4. The wine dataset.} The dataset was retrieved from the UCI repository (see text). It contains three different cultivars, indicated as 1, 2 and 3 in the on line repository and here reported as gray, white and black circles. The top-left panel reports the principal component analysis of this dataset, while all other panels report different runs of random component analysis. The cultivars, particularly those indicated as 2 and 3, partially overlap with both algorithms.}
\end{figure}

\clearpage
\begin{table}[!ht]
\tiny

\begin{adjustwidth}{-2.5in}{0in} 
\begin{tabular}{|c|c|l|c|}
\hline
{\bf \#} & {\bf PDB} & {\bf Ligands} & {\bf Res \AA }\\ 
\hline 
1 & 1BM0 & N/A & 2,5\\ 
\hline 
2 & 1E78 & N/A & 2,6\\ 
\hline 
3 & 1E7A & 2,6-BIS(1-METHYLETHYL)PHENOL (PROPOFOL) & 2,2\\ 
\hline 
4 & 1E7C & MYRISTIC ACID; 2-BROMO-2-CHLORO-1,1,1-TRIFLUOROETHANE & 2,4\\ 
\hline 
5 & 1E7E & DECANOIC ACID & 2,5\\ 
\hline 
6 & 1E7F & LAURIC ACID & 2,43\\ 
\hline 
7 & 1E7G & MYRISTIC ACID & 2,5\\ 
\hline 
8 & 1E7H & PALMITIC ACID & 2,43\\ 
\hline 
9 & 1E7I & STEARIC ACID & 2,7\\ 
\hline 
10 & 1GNI & OLEIC ACID & 2,4\\ 
\hline 
11 & 1GNJ & ARACHIDONIC ACID & 2,6\\ 
\hline 
12 & 1H9Z & R-WARFARIN; MYRISTIC ACID & 2,5\\ 
\hline 
13 & 1HA2 & S-WARFARIN; MYRISTIC ACID & 2,5\\ 
\hline 
14 & 1HK4 & 3,5,3',5'-TETRAIODO-L-THYRONINE; MYRISTIC ACID & 2,4\\ 
\hline 
15 & 1HK5 & 3,5,3',5'-TETRAIODO-L-THYRONINE; MYRISTIC ACID & 2,7\\ 
\hline 
16 & 1N5U & PROTOPORPHYRIN IX CONTAINING FE (HEME); MYRISTIC ACID & 1,9\\ 
\hline 
17 & 1O9X & PROTOPORPHYRIN IX CONTAINING FE (HEME); MYRISTIC ACID & 3,2\\ 
\hline 
18 & 1UOR & N/A & 2,8\\ 
\hline 
19 & 2BX8 & AZAPROPAZONE & 2,7\\ 
\hline 
20 & 2BXB & 4-BUTYL-1-(4-HYDROXYPHENYL)-2-PHENYLPYRAZOLIDINE- 3,5-DIONE (OXYPHENBUTAZONE) & 3,2\\ 
\hline 
21 & 2BXC & 4-BUTYL-1,2-DIPHENYL-PYRAZOLIDINE-3,5-DIONE & 3,1\\ 
\hline 
22 & 2BXD & R-WARFARIN & 3,05\\ 
\hline 
23 & 2BXF & 7-CHLORO-1-METHYL-5-PHENYL-1,3-DIHYDRO-2H- 1,4-BENZODIAZEPIN-2-ONE & 2,95\\ 
\hline 
24 & 2BXG & 2-(4-ISOBUTYLPHENYL)PROPIONIC ACID (IBUPROFEN) & 2,7\\ 
\hline 
25 & 2BXI & AZAPROPAZONE; MYRISTIC ACID & 2,5\\ 
\hline 
26 & 2BXK & INDOMETHACIN; AZAPROPAZONE; MYRISTIC ACID & 2,4\\ 
\hline 
27 & 2BXL & 2-HYDROXY-3,5-DIIODO-BENZOIC ACID; MYRISTIC ACID & 2,6\\ 
\hline 
28 & 2BXM & INDOMETHACIN; MYRISTIC ACID & 2,5\\ 
\hline 
29 & 2BXN & 3-[5-[(3-CARBOXY-2,4,6-TRIIODO-PHENYL)CARBAMOYL]PENTANOYLAMINO]- 2,4,6-TRIIODO-BENZOIC ACID; MYRISTIC ACID & 2,65\\ 
\hline 
30 & 2BXO & 4-BUTYL-1-(4-HYDROXYPHENYL)-2-PHENYLPYRAZOLIDINE- 3,5-DIONE (OXYPHENBUTAZONE); MYRISTIC ACID & 2,6\\ 
\hline 
31 & 2BXP & 4-BUTYL-1,2-DIPHENYL-PYRAZOLIDINE-3,5-DIONE; MYRISTIC ACID & 2,3\\ 
\hline 
32 & 2BXQ & INDOMETHACIN; 4-BUTYL-1,2-DIPHENYL-PYRAZOLIDINE-3,5-DIONE; MYRISTIC ACID & 2,6\\ 
\hline 
33 & 2I2Z & MYRISTIC ACID; 2-HYDROXYBENZOIC ACID (SALICYLIC ACID) & 2,7\\ 
\hline 
34 & 2I30 & MYRISTIC ACID; 2-HYDROXYBENZOIC ACID (SALICYLIC ACID) & 2,9\\ 
\hline 
35 & 2XSI & DANSYL-L-GLUTAMATE; MYRISTIC ACID & 2,7\\ 
\hline 
36 & 2XVV & DANSYL-L-ASPARAGINE; MYRISTIC ACID & 2,4\\ 
\hline 
37 & 2XVW & DANSYL-L-ARGININE; MYRISTIC ACID & 2,65\\ 
\hline 
38 & 3B9L & 3'-AZIDO-3'-DEOXYTHYMIDINE (AZIDOTHYMIDINE); MYRISTIC ACID & 2,6\\ 
\hline 
39 & 3B9M & 3'-AZIDO-3'-DEOXYTHYMIDINE (AZIDOTHYMIDINE); MYRISTIC ACID;  2-HYDROXYBENZOIC ACID (SALICYLIC ACID) & 2,7\\ 
\hline 
40 & 3CX9 & (2S)-3-{[(R)-(2-AMINOETHOXY)(HYDROXY)PHOSPHORYL]OXY}- 2-HYDROXYPROPYL HEXADECANOATE; MYRISTIC ACID & 2,8\\ 
\hline 
41 & 3JRY & SULFATE ION & 2,3\\ 
\hline 
42 & 3LU6 & [(1R,2R)-2-{[(5-FLUORO-1H-INDOL-2-YL)CARBONYL]AMINO}- 2,3-DIHYDRO-1H-INDEN-1-YL]ACETIC ACID & 2,7\\ 
\hline 
43 & 3LU7 & 4-[(1R,2R)-2-{[(5-FLUORO-1H-INDOL-2-YL)CARBONYL]AMINO}- 2,3-DIHYDRO-1H-INDEN-1-YL]BUTANOIC ACID; PHOSPHATE ION & 2,8\\ 
\hline 
44 & 3SQJ & MYRISTIC ACID & 2,05\\ 
\hline 
45 & 3UIV & MYRISTIC ACID; (3S,5S,7S)-TRICYCLO[3.3.1.1~3,7~]DECAN-1- AMINE (AMANTADINE) & 2,2\\ 
\hline 
46 & 4BKE & PALMITIC ACID & 2,35\\ 
\hline 
47 & 4G03 & N/A & 2,22\\ 
\hline 
48 & 4G04 & N/A & 2,3\\ 
\hline 
49 & 4IW2 & ALPHA-D-GLUCOSE; D-GLUCOSE IN LINEAR FORM; PHOSPHATE ION & 2,41\\ 
\hline 
50 & 4K2C & N/A & 3,23\\ 
\hline 
51 & 4L8U & (2S)-2-[1-AMINO-8-(HYDROXYMETHYL)-9-OXO-9,11- DIHYDROINDOLIZINO[1,2-B]QUINOLIN-7-YL]-2- HYDROXYBUTANOIC ACID; MYRISTIC ACID & 2,01\\ 
\hline 
52 & 4L9K & (2S)-2-HYDROXY-2-[8-(HYDROXYMETHYL)-9-OXO- 9,11-DIHYDROINDOLIZINO[1,2-B]QUINOLIN-7-YL]BUTANOIC ACID & 2,4\\ 
\hline 
53 & 4L9Q & TENIPOSIDE & 2,7\\ 
\hline 
54 & 4LA0 & R-BICALUTAMIDE & 2,4\\ 
\hline 
55 & 4LB2 & IDARUBICIN & 2,8\\ 
\hline 
56 & 4LB9 & ETOPOSIDE; MYRISTIC ACID & 2,7\\ 
\hline 
57 & 4Z69 & 2-[2,6-DICHLOROPHENYL)AMINO]BENZENEACETIC ACID (DICLOFENAC); PALMITIC ACID; PENTADECANOIC ACID & 2,19\\ 
\hline 
58 & 5IFO & MYRISTIC ACID; RUTHENIUM ION & 3,2\\ 
\hline 

\end{tabular}
\caption*{{\bf Supplementary Table 1.} Ligands, resolution and literature references of the HSA dataset.}
\end{adjustwidth}
\end{table}

\clearpage

\begin{table}[!ht]
\scriptsize
\centering
\begin{adjustwidth}{-0.75in}{0in}
\begin{tabular}{|c|c|c|c|c|c|}
\hline
 \# & PDB & bound fatty acids & Qh tree & RMSD Tree & RCA cluster    \\ 
\hline
1 & 1BM0 & no & A & A & A    \\ 
\hline
2 & 1E78 & no & A & A & A    \\ 
\hline
3 & 1E7A & no & A & A & A    \\ 
\hline
4 & 1E7C & yes & B & B & B    \\ 
\hline
5 & 1E7E & yes & B & B & B    \\ 
\hline
6 & 1E7F & yes & B & B & B    \\ 
\hline
7 & 1E7G & yes & B & B & B    \\ 
\hline
8 & 1E7H & yes & B & B & B    \\ 
\hline
9 & 1E7I & yes & B & B & B    \\ 
\hline
10 & 1GNI & yes & B & B & B    \\ 
\hline
11 & 1GNJ & yes & B & B & B    \\ 
\hline
12 & 1H9Z & yes & B & B & B    \\ 
\hline
13 & 1HA2 & yes & B & B & B    \\ 
\hline
14 & 1HK4 & yes & B & B & B    \\ 
\hline
15 & 1HK5 & yes & B & B & B    \\ 
\hline
16 & 1N5U & yes & B & B & B    \\ 
\hline
17 & 1O9X & yes & B & B & B    \\ 
\hline
18 & 1UOR & no & A & A & A    \\ 
\hline
19 & 2BX8 & no & A & A & A    \\ 
\hline
20 & 2BXB & no & A & A & A    \\ 
\hline
21 & 2BXC & no & A & A & A    \\ 
\hline
22 & 2BXD & no & A & A & A    \\ 
\hline
23 & 2BXF & no & A & A & A    \\ 
\hline
24 & 2BXG & no & A & A & A    \\ 
\hline
25 & 2BXI & yes & B & B & B    \\ 
\hline
26 & 2BXK & yes & B & B & B    \\ 
\hline
27 & 2BXL & yes & B & B & B    \\ 
\hline
28 & 2BXM & yes & B & B & B    \\ 
\hline
29 & 2BXN & yes & B & B & B    \\ 
\hline
30 & 2BXO & yes & B & B & B    \\ 
\hline
31 & 2BXP & yes & B & B & B    \\ 
\hline
32 & 2BXQ & yes & B & B & B    \\ 
\hline
33 & 2I2Z & yes & B & B & B    \\ 
\hline
34 & 2I30 & yes & B & B & B    \\ 
\hline
35 & 2XSI & yes & B & B & B    \\ 
\hline
36 & 2XVV & yes & B & B & B    \\ 
\hline
37 & 2XVW & yes & B & B & B    \\ 
\hline
38 & 3B9L & yes & B & B & B    \\ 
\hline
39 & 3B9M & yes & B & B & B    \\ 
\hline
40 & 3CX9 & yes & B & B & B    \\ 
\hline
41 & 3JRY & no & A & A & A    \\ 
\hline
42 & 3LU6 & no & A & A & A    \\ 
\hline
43 & 3LU7 & no & A & A & A    \\ 
\hline
44 & 3SQJ & yes & B & B & B    \\ 
\hline
45 & 3UIV & yes & B & B & B    \\ 
\hline
46 & 4BKE & yes & B & B & B    \\ 
\hline
47 & 4G03 & no & A & A & A    \\ 
\hline
48 & 4G04 & no & A & A & A    \\ 
\hline
49 & 4IW2 & no & A & A & A    \\ 
\hline
50 & 4K2C & no & A & A & A    \\ 
\hline
51 & 4L8U & yes & B & B & B    \\ 
\hline
52 & 4L9K & no & A & A & A    \\ 
\hline
53 & 4L9Q & no & A & A & A    \\ 
\hline
54 & 4LA0 & no & A & A & A    \\ 
\hline
55 & 4LB2 & no & A & A & A    \\ 
\hline
56 & 4LB9 & yes & B & B & B    \\ 
\hline
57 & 4Z69 & yes & B & B & B    \\ 
\hline
58 & 5IFO & yes & B & B & B  \\
\hline
\end{tabular}
\caption*{{\bf Supplementary Table 2.} HSA dataset clusters.}
\end{adjustwidth}
\end{table}

\clearpage
\subsection*{Numerics}

The Python functions below~perform the principal component analysis (PCA)  and random component analysis (RCA).
\newline
\newline
\begin{verbatim}
#import pandas as pd #uncomment if needed
import numpy as np 
import matplotlib.pyplot as plt 
import pylab as pl
#from scipy import stats #uncomment if needed


def GOE(N):
    """
    Returns a random N x N matrix  of the GOE ensemble. !
    """
    m = np.random.standard_normal((N, N))
    m = (m + np.transpose(m))/2
    return m


#This function implements the classical PCA
def PCA(data):
    """Returns the dataset projected onto the principal basis system"""
    data -= data.mean(axis=0)
    R_matrix = np.corrcoef(data.T)
    eig_vals, eig_vecs = np.linalg.eig(R_matrix)
    #sort eigenvalues and eigenvectors in decreasing order
    index = np.argsort(eig_vals)[::-1]
    eig_vecs = eig_vecs[:, index]
    #
    #performs a test on the lengh of the eigenvectors
    for eigvector in eig_vecs:
        np.testing.assert_array_almost_equal(1.0, np.linalg.norm(eigvector))
    #if the test is not passed, Numpy returns an error message!
    return np.dot(eig_vecs.T, data.T).T, eig_vals


#This function implements the random component analysis
def RCA(data, N):
    """Returns the dataset projected onto a random basis system"""
    data -= data.mean(axis=0)
    R_matrix = GOE(N)
    eig_vals, eig_vecs = np.linalg.eig(R_matrix)
    #sort eigenvalues and eigenvectors in decreasing order
    index = np.argsort(eig_vals)[::-1]
    eig_vecs = eig_vecs[:, index]
    #
    #performs a test on the lengh of the eigenvectors
    for eigvector in eig_vecs:
        np.testing.assert_array_almost_equal(1.0, np.linalg.norm(eigvector))
    #if the test is not passed, Numpy returns an error message!
    return np.dot(eig_vecs.T, data.T).T, eig_vals


def RCA1(data, M):
    """Returns the dataset projected onto a random basis system.
       The random matrix should be determined externally.
    """
    data -= data.mean(axis=0)
    R_matrix = M
    eig_vals, eig_vecs = np.linalg.eig(R_matrix)
    #sort eigenvalues and eigenvectors in decreasing order
    index = np.argsort(eig_vals)[::-1]
    eig_vecs = eig_vecs[:, index]
    #
    #performs a test on the lengh of the eigenvectors
    for eigvector in eig_vecs:
        np.testing.assert_array_almost_equal(1.0, np.linalg.norm(eigvector))
    #if the test is not passed, Numpy returns an error message!
    return np.dot(eig_vecs.T, data.T).T, eig_vals

\end{verbatim}

\end{document}